\documentclass[aps, prb, graphicx, 12pt, reprint, floatfix]{revtex4-1}
\usepackage[utf8]{inputenc}

\usepackage{graphicx}
\usepackage{amsmath}
\usepackage{amssymb}
\usepackage[hidelinks]{hyperref}
\usepackage{multirow}
\usepackage{soul}
\usepackage{xcolor}
\usepackage[font=small,labelfont=bf]{caption} 

\usepackage[version=3]{mhchem} 

\newcommand{\fmax}{f_{\mathrm{max}}}
\newcommand{\imax}{i_\mathrm{max}}
\newcommand{\fscale}{\alpha_\mathrm{s}}

\begin{document}

\title{Scaled and Dynamic Optimizations of Nudged Elastic Bands}
\author{Per Lindgren}
\author{Georg Kastlunger}
\author{Andrew A. Peterson}
\email{andrew\_peterson@brown.edu}
\affiliation{School of Engineering, Brown University, Providence, Rhode Island, 02912, USA}

\begin{abstract}
\noindent
We present a modified nudged elastic band routine that can reduce the number of force calls by more than 50\% for bands with non-uniform convergence. The method, which we call ``dyNEB", dynamically and selectively optimizes states based on the perpendicular PES-derived forces and parallel spring forces acting on that region of the band. The convergence criteria are scaled to focus on the region of interest, \textit{i.e.}, the saddle point, while maintaining continuity of the band and avoiding truncation. We show that this method works well for solid state reaction barriers---non-electrochemical in general and electrochemical in particular---and that the number of force calls can be significantly reduced without loss of resolution at the saddle point.
\end{abstract}

\maketitle

Reaction barrier estimates are pivotal in computational studies of chemical systems; reaction rates in homogeneous and heterogeneous catalysis, alloy stability~\cite{Vej-Hansen2016} and current densities in electrochemistry~\cite{Skulason2007, Skulason2010, Head-Gordon_2016, He2017, Singh2017}, to name a few, all depend upon the height of the transition state separating two thermodynamically stable states. Kinetic information from \textit{ab initio} calculations can be used to calculate rate constants via transition state theory~\cite{Eyring1935} and connect microscopic quantities to macroscopic observables. This information is contained on many-dimensional potential energy surfaces (PES)---the central theme in computational chemistry. Here, thermodynamically stable states are located in local minima, and the transition from one state to another depends upon the height of the transition state region separating the two basins. Transition states are first-order saddle points---regions on the PES with convex and concave curvature---and represent the lowest energy region connecting two minima. Saddle point calculations are typically categorized based on the information needed to initialize the calculation~\cite{Jensen2017}. Some methods, notably the Dimer method~\cite{Henkelman1999}, require only local information, while chain-of-states methods require two fixed endpoints. In the former, saddle points are located based on the lowest curvature out of the local minimum. While computationally efficient, these methods require statistical sampling to ensure that the relevant saddle point---and not just any saddle point---is found~\cite{Henkelman1999}. 

The nudged elastic band (NEB)~\cite{NEB_chapter, Henkelman2000, CINEB} method is one of the most widely used algorithms for (first-order) saddle point calculations. This chain-of-states method connects two thermodynamically stable states by intermediate states and Hookean springs. These springs allow the intermediate states to traverse high-energy regions of the PES. The nudging force projection, after which the method is named, only includes spring forces parallel to the path and PES--derived forces perpendicular to the reaction pathway (Eq.~\ref{eq:nudging})~\cite{Henkelman2000},

\begin{equation}
\mathbf{F}_i = \mathbf{F}_i^s \big|_\parallel - \nabla V(\mathbf{R}_i) \big|_\perp.
 \label{eq:nudging}
\end{equation}

The chain-of-states is conventionally initialized by linear interpolation between the initial and final states or by more sophisticated interpolation schemes~\cite{Smidstrup2014}, and the band is subsequently optimized to locate the minimum energy pathway (MEP) connecting the two basins. Although rigorous and widely used throughout computational chemistry, the NEB method is often hampered by its significant computational cost. This is due in part to size (all states must be simultaneously optimized) and the dimensionality of the optimization problem. Many methods have been proposed to alleviate the computational cost, ranging from adaptive bands~\cite{Maragakis2002} and automated methods~\cite{Kolsbjerg2016} to reflection symmetry operations~\cite{Mathiesen2019} and machine-learning algorithms~\cite{Peterson2016, Koistinen2017, Garrido2019}. 

Recent efforts in the electrocatalysis community have focused on developing constant-potential density functional theory (DFT)~\cite{Hohenberg_Kohn_DFT_1964, Kohn_Sham_DFT_1965} methods. These electronically grand canonical methods~\cite{Alavi_JCP_2001, Sugino_PRB_2006, Neurock_2006, Jinnouchi2008, Letchworth-Weaver2012, Head-Gordon_2016,Sundararaman2017, Kastlunger2018, Bouzid2018, Melander2019} allow the number of electrons in the system to fluctuate along the reaction trajectory in order to keep the applied potential (work function) constant. Such schemes are required when calculating electrochemical reaction barriers, since finite-sized unit cells introduce an artificial potential bias along the reaction trajectory in constant-charge DFT~\cite{Rossmeisl2008, Skulason2010}. The computational cost of NEB calculations can grow manifold when coupled with these semi-grand canonical methods, since the geometry of the band and the work function of each state along the reaction trajectory must be simultaneously optimized. In this emerging field, there is need for a computationally efficient approach to reaction barrier calculations. 

Part of the allure of the NEB method is that it is embarrassingly parallel; computational resources can be efficiently partitioned to calculate all interior states simultaneously. This is certainly true for bands with uniform convergence. However, as we will show, convergence of bands is often highly non-uniform, and significant computational resources are often spent calculating states that are well below the convergence criterion. This issue is particularly severe for semi-grand canonical saddle point searches. Moreover, the efficiency of parallel schemes not only hinges on uniform convergence of interior states, but also on homogeneous computer architectures.

In this Letter, we introduce a methodology to improve saddle point convergence for non-uniform MEPs. Moreover, we demonstrate that an electronically grand canonical treatment makes convergence of the MEP inherently non-uniform, since charge equilibration is particularly important at or around the saddle point~\cite{Kastlunger2018}. This makes parallel NEB schemes unsuitable, since a large fraction of computational resources will idle for significant parts of the optimization while a small fraction is used to optimize the region of interest, \textit{i.e.}, the saddle point. We present a serial NEB implementation that \textit{dynamically} and \textit{selectively} optimizes the reaction pathway in NEB calculations. While this scheme is applicable to any NEB calculation, we note that it is particularly well-suited for semi-grand canonical saddle point searches.

\begin{figure*}[t]
 \includegraphics[width=0.99\textwidth]{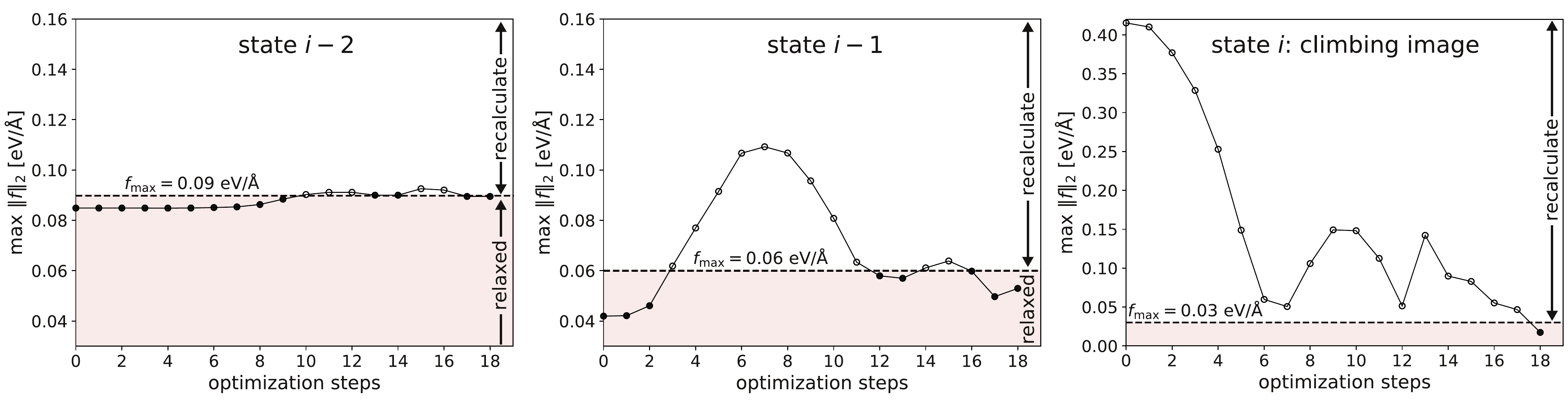}
 \caption{Dynamic optimization for the associative desorption of hydrogen on Au(111). Three interior states are shown: the climbing image, its nearest neighbor and a state two springs removed. The dramatic geometry change of the climbing image pulls the neighboring states out of convergence. Open and filled circles represent recalculated and converged states, respectively.}
\label{fig:dyn}
\end{figure*}

Convergence of chains-of-states is often highly non-uniform; states in low-energy regions of the PES (in closer vicinity to local minima) generally converge faster than those close to the saddle point. Moreover, in all but a few applications, the important result of an NEB calculation is the saddle point. The interior states before and after the saddle point contribute to mapping out a reasonable minimum energy pathway and provide tangent estimates for the nudging force projection (Eq.~\ref{eq:nudging}), but are typically not used in subsequent analyses. Thus, the computational effort can be significantly reduced when optimizing states to different convergence criteria. We do this by scaling the convergence criteria along the reaction trajectory and dynamically optimizing the states. A state is not recalculated if the forces acting on that state are below the convergence criterion; if the forces rise above the convergence criterion---due to spring adjustments between the state and its neighbors---the state is recalculated, hence \textit{dynamic}. This convergence check is performed after each optimization step. Fig.~\ref{fig:dyn} shows an example of dynamic optimizations for three states in the associative desorption reaction of hydrogen on Au(111) when the climbing image~\cite{CINEB} implementation of NEB is invoked. For simplicity, we consider a simple convergence scaling based on state indices, $\fmax^i = \fmax^{\imax} \cdot \left( 1 + | \imax - i|  \cdot \fscale \right)$, where $\imax$ is the state with the highest potential energy and $\fscale$ is the scaling factor. In Fig.~\ref{fig:dyn}, the $i^\mathrm{th}$ state is the climbing image, and its geometry changes drastically when the state climbs up the gradient of the PES. This changes the parallel spring force, $\mathbf{F}_i^s|_\parallel$, between states $i$ and $i-1$ (although the climbing image does not feel the spring forces), and subsequently pulls state $i-1$ out of convergence. State $i-1$ is then recalculated, and the change in the parallel spring force, $\mathbf{F}_{i-1}^s|_\parallel$, between states $i-1$ and $i-2$ causes state $i-2$ to be recalculated. This implementation is therefore truly dynamic, since all states dynamically respond to any perturbations in the band. Convergence plots for all states of this example are shown in Figs. 2--3 of the supporting information.

\begin{figure*}[t]
 \includegraphics[width=0.95\textwidth]{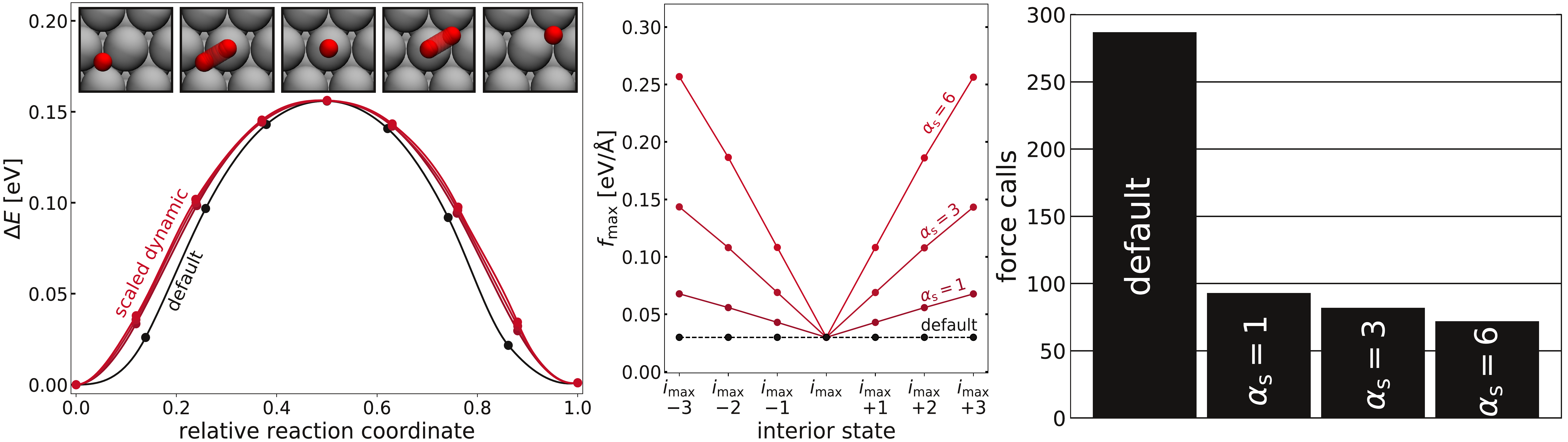}
 \caption{Left: Minimum energy pathway of oxygen diffusion on Pt(111) for the default NEB (black) and dynamic NEB with displacement scaling (red). Center: Convergence criteria for the scaled dynamic implementation. The convergence criterion for the state with the highest energy is $\fmax^{\imax}=0.03$ eV/\AA{}.
 \label{fig:O_fcc}}
 \end{figure*}

The choice of convergence scaling is in principle arbitrary, but a very high scaling factor could disrupt the pathway (see Fig. 4 of the supporting information). This highlights the importance of dynamic optimizations, since ``frozen" states can be perturbed away from the MEP if the geometries of their neighboring states change. This is particularly important when using the climbing-image~\cite{CINEB} implementation of NEB, since the geometry (and number of electrons in electronically grand canonical methods) changes significantly in this region of the PES. We find that high scaling factors only introduce modest changes in the number of force calls compared to a more conservative convergence scaling (Fig.~\ref{fig:O_fcc} and table~\ref{tab:Tafel}), and could disrupt the pathway and delay convergence. Thus, a low scaling factor is recommended. Here, we choose to scale the convergence criteria based on a displacement metric, 

\begin{equation}
\fmax^i = \fmax^{\imax} \cdot \left( 1 + \sqrt{\sum_{j=0}^N \left(\mathbf{R}_{\imax, j} - \mathbf{R}_{i, j} \right)^2}  \cdot \fscale \right),
\label{eq:pos_scale}
\end{equation}

\noindent
where $i$ loops over all interior states, $\imax$ is the index of the interior state with the highest potential energy, $\fmax^{\imax}$ is the convergence criterion given to the optimizer, $\mathbf{R}_i$ is the position matrix of the $i^\mathrm{th}$ state, $j$ is the atom index, $N$ is the number of atoms and $\fscale$ is the scaling factor. The advantage of this implementation is that it is independent of sampling density along the band. We emphasize that this convergence scaling is dynamic; $\fscale$ is assigned \textit{a priori}, but $\imax$ and $\mathbf{R}_{i}$ are updated after each optimization step. We have implemented this scaled and dynamic optimization method of nudged elastic bands in the Atomic Simulation Environment (ASE)~\cite{ase-paper}.

As a simple example, we consider oxygen diffusion on Pt(111). Fig.~\ref{fig:O_fcc} shows the MEP for the default NEB implementation (black) and ``dyNEB" with different convergence scaling factors, as calculated with Effective Medium Theory~\cite{Jacobsen1996}. Here, the pathway leading to the saddle point differs between the two methods; the tails of the bands converge quickly, and are thus unnecessarily recalculated in the default implementation. However, the saddle point geometry and energy are identical in the two methods, and the reduction in force calls is significant; the highest scaling ($\fscale = 6$) of the convergence criteria reduces the number of force calls by 75\%. We note that a dynamic relaxation without convergence scaling ($\fscale = 0$) reduces the number of force calls by 59\% for this particular reaction barrier.

We use the method outlined above the calculate (i) non-electrochemical diffusion barriers and (ii) electrochemical reaction barriers at constant applied potential with the Solvated Jellium (SJ) method~\cite{Kastlunger2018}, and show that the number of force calls in NEB calculations can be significantly reduced without loss of resolution at the saddle point. However, the method is equally applicable to molecular saddle point searches. Each benchmark calculation is started from linear interpolation, and the climbing-image method is only invoked after the band has converged. All computational details are listed in the supporting information.

\begin{figure}[t]
 \includegraphics[width=0.49\textwidth]{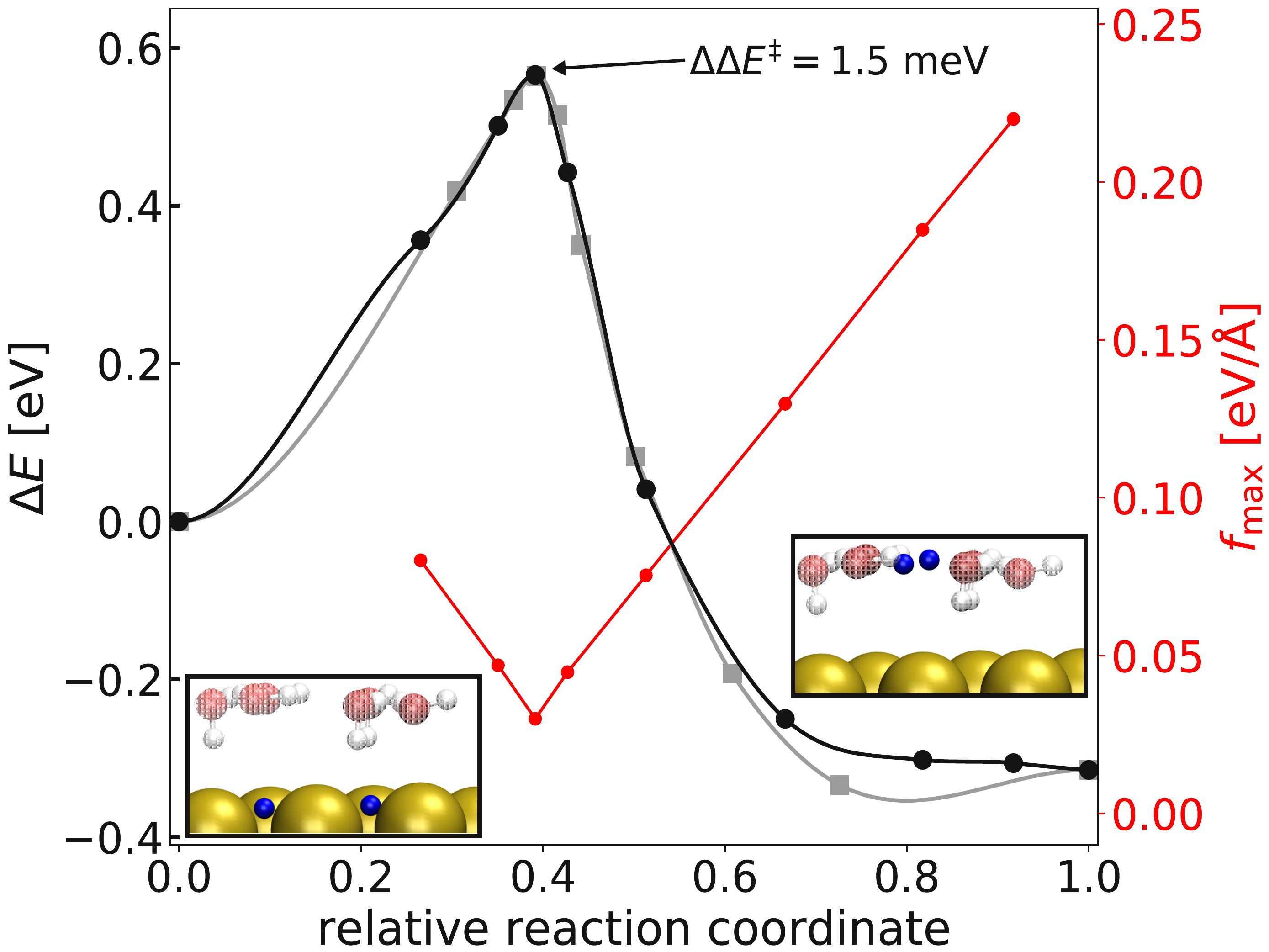}
\caption{Reaction pathway for the Tafel reaction on Au(111) for $\fscale=2$ (black circles) and the default (gray squares) NEB method. The convergence scaling is shown in red. The energy and geometry of the saddle point are identical in both implementations.}
\label{fig:Tafel}
 \end{figure}

The non-electrochemical associative desorption of hydrogen (Tafel) of the hydrogen evolution reaction (HER) can be calculated without electronically grand canonical methods, since the charge transfer between the initial and final states (and any states along the reaction trajectory) is negligible. Fig.~\ref{fig:Tafel} shows the converged pathway on Au(111) for ``dyNEB" with $\fscale=2$ in black and the default implementation in gray. The implementations differ only in the convergence of states away from the saddle point; states far away from the saddle point converge faster and to a higher convergence criterion, and thus require less force calls. Note that the convergence scaling in Eq.~\ref{eq:pos_scale} captures the asymmetry of the reaction pathway. 

However, the geometry and energy of the saddle point are identical in both implementations. As shown in table~\ref{tab:Tafel}, this dynamic and selective optimization method rapidly locates the saddle point; ``dyNEB" with $\fscale=2$ reduces the total number of force calls and CPU time by 58\% and 52\%, respectively, compared to a default parallel implementation. Thus, the slightly higher CPU time--per--force call of ``dyNEB", which stems from the serial implementation, is offset by a significant reduction in the number of force calls. Moreover, the reduction in force calls is more pronounced when the climbing-image method is invoked; $\fscale=2$ decreases the number of force calls by 69\% compared to the default methodology.

\begin{table}
\begin{tabular}{l  c | c | c  | c}
Method & & $\Delta E^{\ddag}$ (eV) & Force calls & Normalized \\ & & & & CPU time \\
\hline
\hline
\multirow{2}{1.2cm}{Default \\ parallel} & no climbing & --- & 1,960  &  1.0 \\
                          &  climbing & 0.564 & 224  &  1.0   \\
\hline
\multirow{2}{1.2cm}{$\fscale=1$} & no climbing & --- &  991  &   0.56   \\
                       & climbing & 0.563 & 93 &  0.49  \\
\hline
\multirow{2}{1.2cm}{$\fscale=2$} & no climbing & --- & 857 & 0.49 \\
                       & climbing & 0.566 & 70  &  0.39  \\
\hline
\end{tabular}
\caption{Force calls and normalized CPU time for a saddle point calculation of the Tafel reaction (Fig.~\ref{fig:Tafel}) with eight interior states.}
\label{tab:Tafel}
\end{table}

\begin{figure*}[t]
 \includegraphics[width=0.99\textwidth]{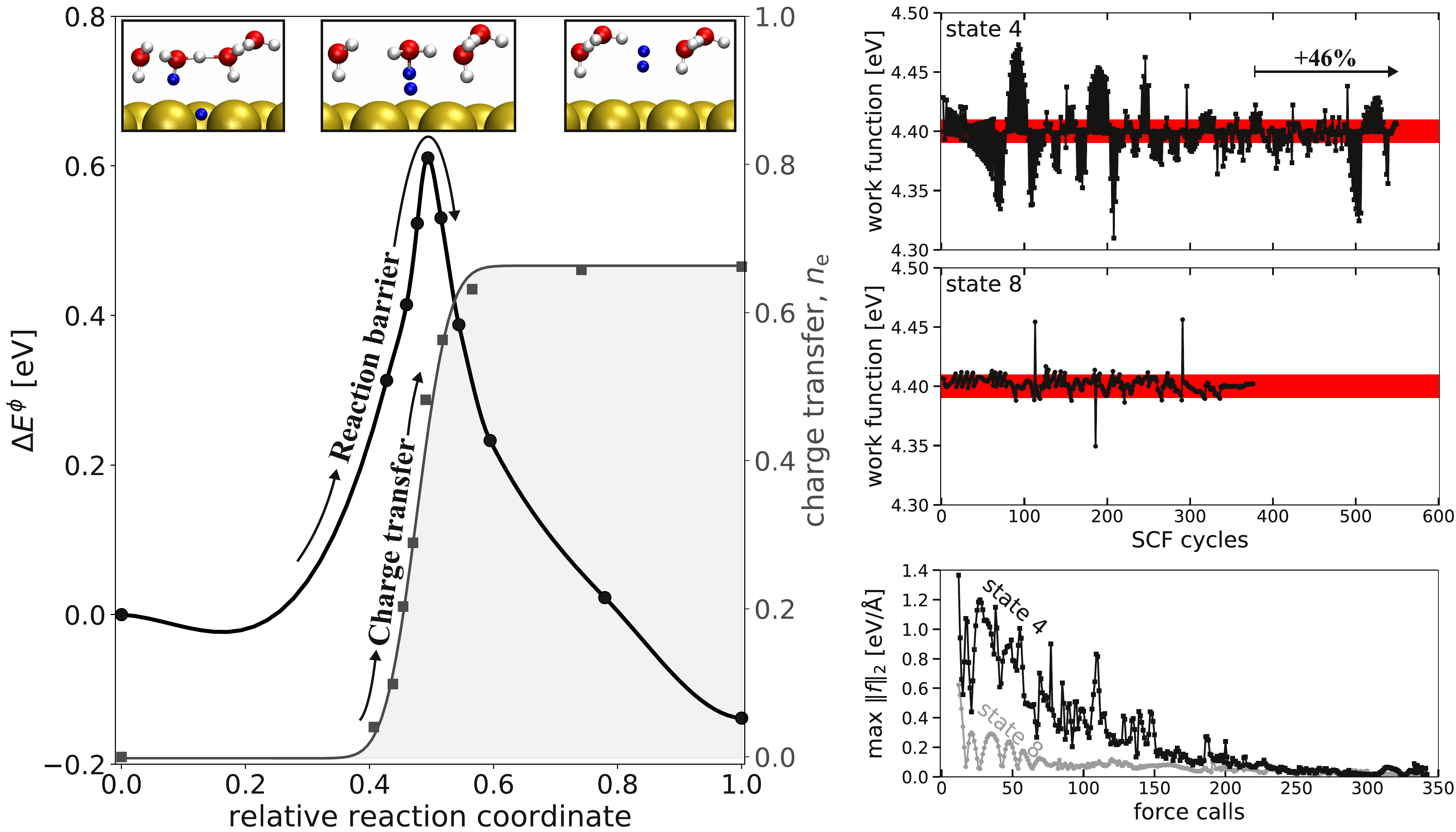}
\caption{Left: Minimum energy pathway (circles) and charge transfer profile (squares) of the Heyrovsky reaction of HER on Au(111) at the equilibrium potential. Right: Work function as a function of SCF cycles and forces as a function of number of force calls for states close to the saddle point (black squares) and close to the final state (gray circles). The red region represents the work function tolerance ($\pm0.01$ eV) in the constant-potential scheme.}
\label{fig:Hey}
 \end{figure*}

The non-uniformity of convergence of states is exacerbated for semi-grand canonical treatments of electrochemical reaction barriers; not only do some interior states converge faster than others, but the iterative work function calculation is more involving for states close to the saddle point. We show this for the Heyrovsky reaction ($\ast \mathrm{H} \ +$ \ce{H+ + e-} $\rightarrow \  \ast \ +$ \ce{H2}) of HER on Au(111), as calculated with the Solvated Jellium (SJ) method~\cite{Kastlunger2018}. The left panel of Fig.~\ref{fig:Hey} shows the minimum energy pathway (circles) and charge transfer profile (squares). Note that the spring constants are adapted to increase the resolution around the saddle point and decrease the sampling density in the two diffusion regions. The fractional charge transfer between the endstates is a consequence of hybridization between the solvated proton and the electrode, as noted in a recent study~\cite{Chen2018}. The adiabatic nature of this calculation manifests itself in an electron transfer process that is distributed between interior states and has finite width along the reaction coordinate. However, the saddle point is tight and well-defined, and the inflection point of the charge transfer curve coincides with the saddle point. Thus, the iterative work function calculation is particularly important in this region. This is shown in the right panel, where the work function fluctuates more for states around the saddle point (fourth interior state) than for states close to local minima (eighth interior state); the state at the saddle point requires 46\% more self-consistent field (SCF) cycles than the state close to a local minimum. Moreover, the charge equilibration scheme in the SJ method~\cite{Kastlunger2018} is optimized to rapidly counteract any perturbations in the work function, and rarely needs more than two SCF cycles to reach the target potential. Previous studies~\cite{Gauthier2019} suggest that three SCF cycles are required to reach the target potential with Newton's method; this would exacerbate the non-uniformity further. The bottom right panel of Fig.~\ref{fig:Hey} shows the forces acting on a low-energy state (eighth interior state) as well as the forces acting on a state close to the saddle point (fourth interior state). The low-energy state quickly approaches the convergence criterion, but is recalculated in default NEB implementations.

A default parallel implementation of NEB is unsuitable for these types of saddle point calculations, since states in close proximity to the saddle point require more force calls \textit{and} SCF cycles. If resources are parallelized over interior states, then a significant fraction will idle while a small fraction is used to calculate the saddle point. This can be seen in table~\ref{tab:Hey}, where the total number of force calls and CPU time decrease by 51\% and 56\%, respectively, when using ``dyNEB" with $\fscale=3$. Thus, the current implementation not only decreases the number of force calls, but also the CPU time--per--force call for electronically grand canonical saddle point searches.

\begin{table}
\begin{tabular}{l  c | c | c  | c}
Method & & $\Delta E^{\ddag}$ (eV) & Force calls & Normalized \\ & & & & CPU time \\
\hline
\hline
\multirow{2}{1.2cm}{Default \\ parallel} & no climbing & --- & 2,416  & 1.0 \\
                          &  climbing & 0.601 & 304 &  1.0  \\
\hline
\multirow{2}{1.2cm}{$\fscale=2$} & no climbing & --- & 1,829 & 0.77 \\
                       & climbing & 0.607 & 46 & 0.14   \\
\hline
\multirow{2}{1.2cm}{$\fscale=3$} & no climbing & --- & 1,249 & 0.47 \\
                       & climbing & 0.602 & 90 & 0.22   \\
\hline
\end{tabular}
		  \caption{Force calls and normalized CPU time for a semi-grand canonical saddle point calculation of the Heyrovsky reaction (Fig.~\ref{fig:Hey}) with eight interior states.}
\label{tab:Hey}
\end{table}

We have introduced a dynamic and selective implementation of NEB to rapidly locate saddle points. This methodology can significantly reduce the number of force calls in saddle point calculations by carefully monitoring the non-uniform convergence of states along the reaction pathway. The scaled convergence criteria along the trajectory reduce the number of force calls without truncation of the band or loss of resolution at the saddle point. We find that the method works well in the two areas investigated: (i) non-electrochemical and (ii) electrochemical saddle point searches. In the latter category, this method provides a significant advantage to parallel implementations, since the simultaneous optimization of geometry and work function makes convergence of the MEP inherently non-uniform. 

This routine can easily be used as an add-on when the band approaches convergence. That is, the saddle point search can be performed in parallel until some interior states are close to convergence. At that point, the dynamic NEB implementation can be used to avoid unnecessary calculations of converged states. We expect this combination to work efficiently for non-electrochemical and molecular saddle point calculations. For electrochemical saddle points, however, the ``dyNEB" method will always be more efficient than default parallel implementations.

\vspace{5mm}
We gratefully acknowledge funding from the Office of Naval Research, award N00014-16-1-2355, and thank Ask H. Larsen and Esben L. Kolsbjerg for helpful discussions. High-performance computational work was carried out at the Center for Computation \& Visualization (CCV) at Brown University.

\begin{figure*}[t]
 \includegraphics[width=0.9\textwidth]{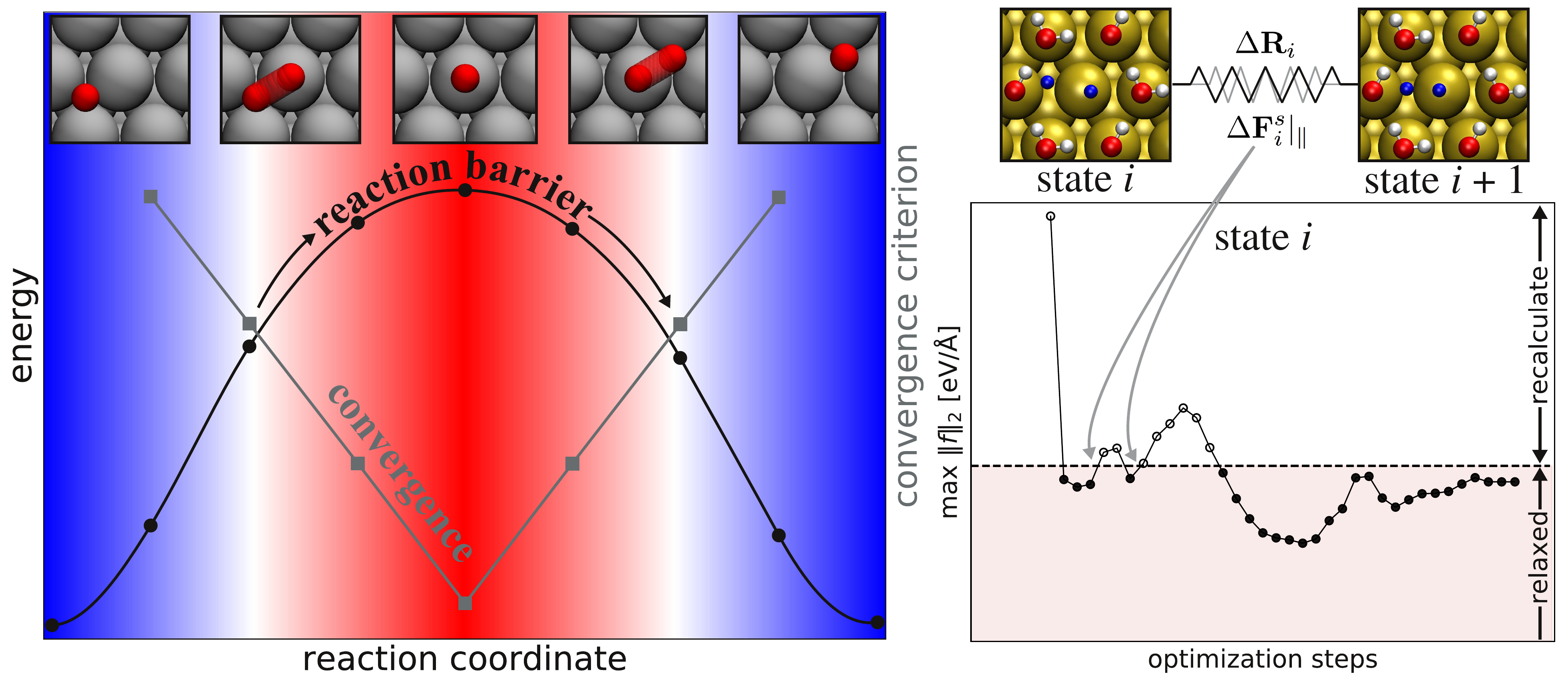}
 \caption{For Table of Contents Only}
 \end{figure*}

\bibliography{neb_paper}
\end{document}